\documentclass{aa}
\usepackage{graphicx}
\usepackage{txfonts}
\usepackage[authoryear]{natbib}
\usepackage{epstopdf}
\begin{document}
\title{Databases and tools for nuclear astrophysics applications}
 \subtitle{BRUSsels Nuclear LIBrary (BRUSLIB), Nuclear Astrophysics Compilation of
REactions II (NACRE II) and Nuclear NETwork GENerator (NETGEN)}

\author{Yi Xu, Stephane Goriely\thanks{Corresponding author. E-mail:
sgoriely@ulb.ac.be}, Alain Jorissen, Guangling Chen, and
Marcel Arnould}
\offprints{Stephane Goriely}
\institute{Institut d'Astronomie et d'Astrophysique, Universit\'e Libre de
Bruxelles, Campus de la Plaine CP 226, 1050, Brussels, Belgium}



\abstract
{An update of a previous description of the BRUSLIB + NACRE package of nuclear data for astrophysics and of the web-based nuclear network generator NETGEN is presented.}{The new version of BRUSLIB contains the latest predictions of a wide variety of nuclear data based on the most recent version of the Brussels-Montreal Skyrme-Hartree-Fock-Bogoliubov model.  The nuclear masses, radii, spin/parities, deformations, single-particle schemes, matter densities, nuclear level densities, E1 strength functions, fission properties, and partition functions are provided for all nuclei lying between the proton and neutron drip lines over the  8 $\leq$ Z $\leq$ 110 range, whose evaluation is based on a unique microscopic model that ensures a good compromise between accuracy, reliability, and feasibility. In addition, these various ingredients are used to calculate about 100000 Hauser-Feshbach neutron-, proton-, $\alpha$-, and $\gamma$-induced reaction rates based on the reaction code TALYS.} {NACRE is superseded by the NACRE II compilation for 15 charged-particle transfer reactions and 19 charged-particle radiative captures on stable targets with mass numbers A $<$ 16. NACRE II features the inclusion of experimental data made available after the publication of NACRE in 1999 and up to 2011. In addition, the extrapolation of the available data to the very low energies of astrophysical relevance is improved through the systematic use of phenomenological potential models. Uncertainties in the rates are also evaluated on this basis.} {Finally, the latest release v10.0 of the web-based tool NETGEN is presented. In addition to the data already used in the previous NETGEN package, it contains in a fully documented form the new BRUSLIB and NACRE II data, as well as new experiment-based radiative neutron capture cross sections.}
{The full new versions of BRUSLIB, NACRE II, and NETGEN are available electronically from the nuclear database at http://www.astro.ulb.ac.be/NuclearData. The nuclear material is presented in an extended tabular form complemented with a variety of graphical interfaces.}						
\authorrunning{Y. Xu, S. Goriely, A. Jorissen, et al.}
\titlerunning{BRUSLIB, NACRE II and NETGEN}
\maketitle

\begin{figure*}
\vskip-0.4cm
\centering
\includegraphics[width=\textwidth]{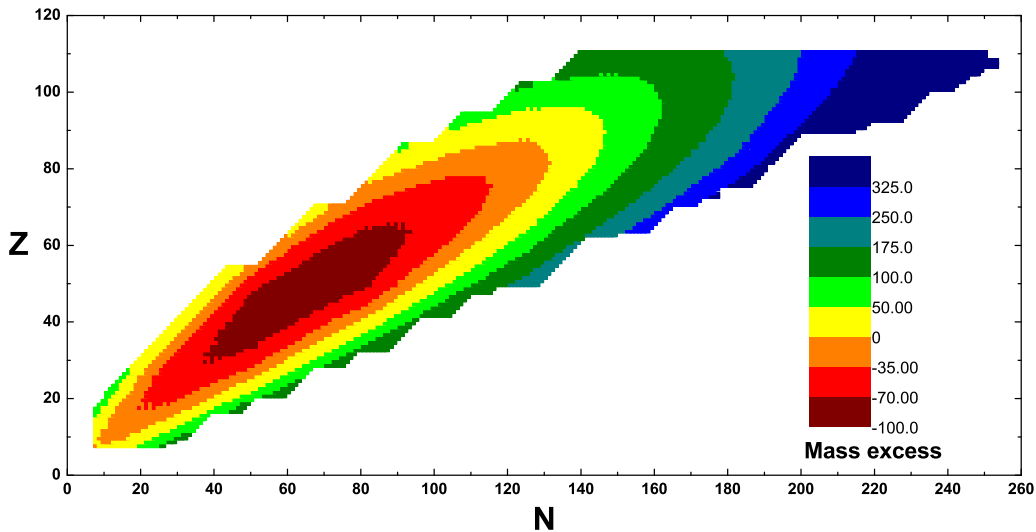}
\vskip-0.5cm
\caption{(Colour online) HFB-21 nuclear mass excesses.}
\label{mcal}
\end{figure*}

\section{Introduction}

Through stellar evolution and stellar or non-stellar nucleosynthesis models,  nuclear astrophysics is in high demand of a huge variety of nuclear data. These include the static properties of nuclei, their spontaneous decay or fission characteristics, and their interactions with a variety of particles and nuclei. In particular, reactions of interest most often concern neutron-, proton-, or $\alpha$-particle induced reactions. Despite of significant efforts, experimental information only covers a minute fraction of the entire required data. It is either insufficient, or even totally non-existent. Massive recourse to theoretical predictions is thus mandatory in many applications.

The availability of compilations that provide an easy access to evaluated and well-documented nuclear data is an essential tool for astrophysics modelling, as is most dramatically illustrated by the impact on this type of research of the pioneering compilation of charged-particle induced reactions by \citet{FCZ67}. This has been the key motivation for the development of a new generation of compilations superseding \citet{FCZ67} and its direct updates (\citet{CF88}, and references therein). The first such second generation compilation is referred to as NACRE (Nuclear Astrophysics Compilation of REactions) \citep{NACRE}. It includes several new important features with respect to its predecessors. As an extension of NACRE, a library of nuclear data referred to as BRUSLIB (BRUSsels LIBrary) has been constructed \citep{AA05}. On top of the NACRE data, it includes a wide variety of nuclear information of interest, and takes full advantage of the internet facility, which allows one to make a huge volume of data available door-to-door in an accurate and usable format. In addition, a nuclear NETwork GENerator NETGEN complements the BRUSLIB package \citep{AA05}. Recently, the Karlsruhe Astrophysical Database of Nucleosynthesis in Stars (KADoNiS) \citep{KADoNiS} and the JINA REACLIB database \citep{REACLIB} have been constructed and are updated regularly.

The aim of this paper is to present an update of BRUSLIB and NETGEN. BRUSLIB now contains new predictions based on microscopic models. As presented in Sect. 2, these include nuclear masses, deformations, single-particle schemes, density distributions, nuclear level densities, partition functions, E1 strength functions, and fission properties. These data are used to calculate a large body of thermonuclear reaction rates based on the Hauser-Feshbach (HF) model. BRUSLIB also contains an update of NACRE, referred to as NACRE II, that is now completed \citep{nacre2pro4,NACRE2}. As discussed in Sect. 3, NACRE II revises 34 NACRE rates based on experimental data published after the NACRE release, and on a new theoretical handling of these data. Section 4 is devoted to the description of the NETGEN update. A summary and some prospects are presented in Sect. 5.

\section{BRUSLIB}

\subsection{Nuclear ground-state properties}

Microscopic models based on realistic nucleon interactions have been used to estimate various nuclear ground-state properties. One of the most promising approaches of this sort is the non-relativistic Hartree-Fock-Bogoliubov (HFB) method, which is based on an effective nucleon-nucleon interaction of the Skyrme type. HFB calculations in which a Skyrme force is fitted to essentially all the mass data \citep{Audi10} can now successfully compete with the most accurate droplet-like formula.

The Skyrme HFB mass model not only seeks optimized fits to the mass data, but it also allows the construction of a universal effective interaction capable of reproducing most of the observables of relevance in nuclear applications. This essential property is obtained by imposing extra physical constraints on the HFB mass
model. In its last version, additional terms are inserted in the effective nucleon-nucleon interaction to better account for different physical effects, among which the pairing interaction, or the properties of neutron matter for the description of the inner crust and the core of neutron stars. Details can be found in \citet{pro1}, \citet{pro2}, \citet{pro3}, \citet{pro4}, \citet{pro5}, \citet{pro6}, \citet{pro7}, \citet{pro8}, \citet{pro9}, \citet{pro10}, and \citet{Goriely10}.

So far, 21 HFB mass tables have been constructed with different parameterizations of the Skyrme effective force. \citet{Goriely10} presented the most recent HFB-21 mass values and related ground-state properties. Tables of 8508 masses containing all nuclides with $8\leq Z \leq 110$ lying between the proton and the neutron drip lines derived from all 21 HFB models can be accessed from BRUSLIB. For each nuclide, the calculated mass is complemented with its nuclear deformation parameters $\beta_{2}$ and $\beta_{4}$, rms charge radius $R_{ch}$, deformation energy $E_{def}$, neutron $S_{n}$ and proton $S_{p}$ separation energies,  beta-decay energy $Q_{bet}$, the difference $M_{err}$ between the experimental and calculated mass excesses as well as the experimental $J_{exp}$ and calculated $J_{th}$ ground-state spins, and the experimental $P_{exp}$ and calculated $P_{th}$ ground-state parities. Figure \ref{mcal} shows the latest HFB-21 mass excesses in the nuclear chart.

By solving the HFB equation, the single-particle spectra, nuclear potential, and charge- and matter density distributions can also be obtained for each nucleus. BRUSLIB provides these results for all 8508 nuclides. These basic nuclear properties provide the essential input required for additional predictions. As examples, the nuclear density distribution is used for microscopic or semi-microscopic optical model potential calculations by folding the target radial matter density with an optical potential in nuclear matter, and the single-particle levels are employed for evaluating the nuclear level density (see Sect. 2.2).

\subsection{Nuclear level densities and partition functions}

\begin{figure*}
\centering
\includegraphics[width=\textwidth]{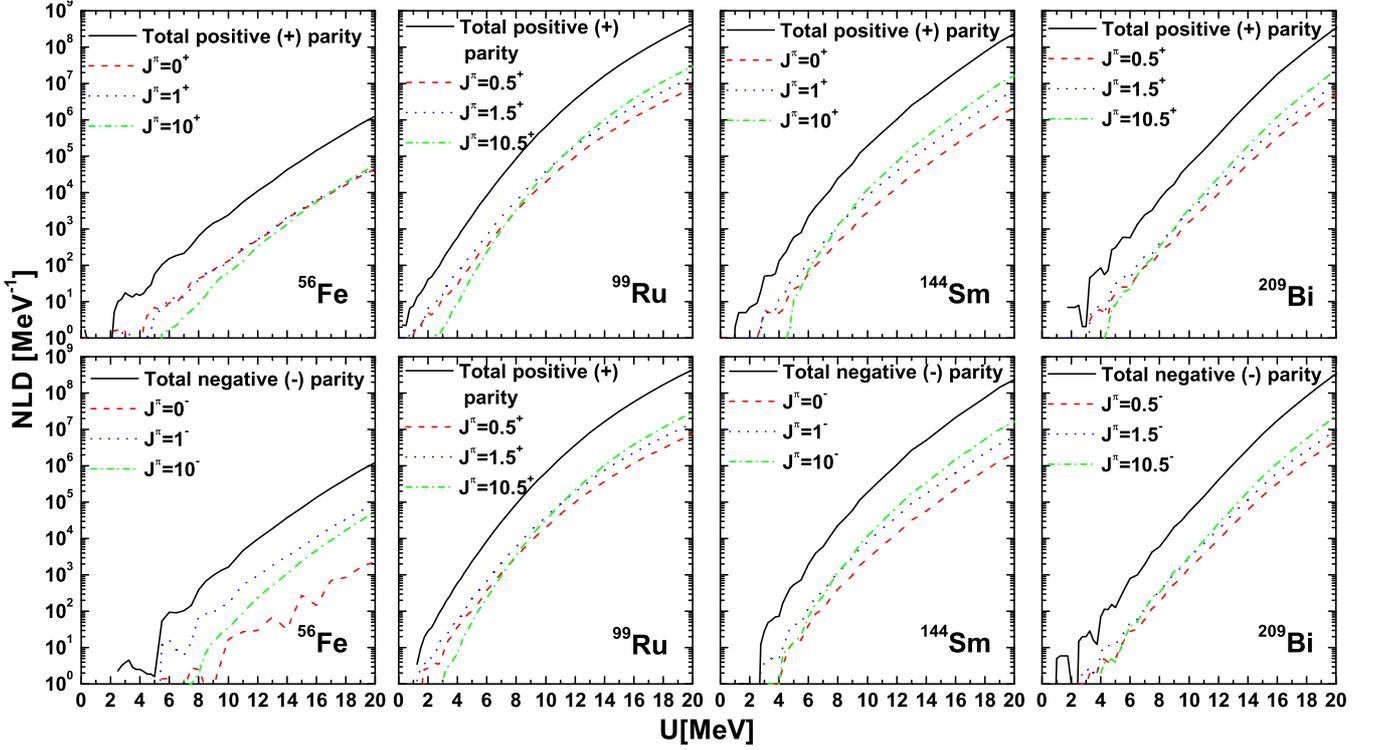}
\vskip-0.2cm
\caption{(Colour online) Spin- and parity-dependent NLDs for $^{56}$Fe, $^{99}$Ru, $^{144}$Sm and $^{209}$Bi predicted by \citet{Goriely08}.}
\label{nld}
\end{figure*}

The nuclear level density (NLD) is an essential ingredient for cross section and reaction rate calculations within the statistical HF model. A microscopic combinatorial approach for estimating the NLDs has been first described by \citet{Hilaire98, Hilaire01}. It has been improved subsequently to include both collective effects and pairing correlations \citep{Hilaire06}.

The model uses the constrained axially symmetric HFB method based on the BSk14 effective Skyrme force \citep{pro8} to construct incoherent particle-hole state densities as functions of the excitation energy, spin projection, and parity. In its latest developments, the combinatorial method improves the description of the collective vibration levels by taking the phonon excitation explicitly into account \citep{Goriely08}.

\begin{figure}
\centering
\includegraphics[width=10.0cm]{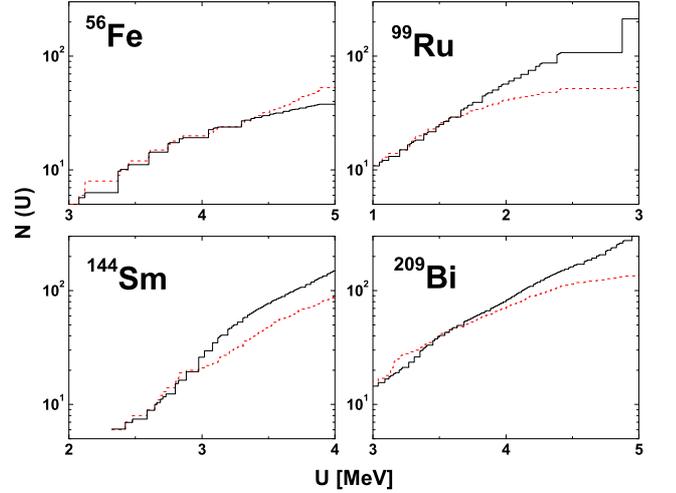}
\vskip-0.2cm
\caption{(Colour online) Comparison of the cumulative numbers of energy levels derived from experiment (solid lines) \citep{RIPL3} and those predicted (dashed line) by the microscopic NLD model of \citet{Goriely08} for $^{56}$Fe, $^{99}$Ru, $^{144}$Sm, and $^{209}$Bi.}
 \label{level}
\end{figure}

BRUSLIB provides the spin- and parity-dependent NLDs from \citet{Goriely08} for 8508 nuclides with $8\leq Z\leq 110$ lying between the proton and the neutron drip lines, for excitation energies $U$ up to 200 MeV, and for spins up to J = 49 for even-$A$ nuclei or 99/2 for odd-$A$ nuclei. The nuclear temperature, cumulative and total numbers of levels are also included. No simple analytical fits to the tabulated NLDs are given to avoid losing the specific microscopic characteristics of the model. A re-normalization procedure of the NLD on experimental data is required in many instances, and in particular for nuclear data evaluation or for an accurate and reliable estimate of reaction cross sections. More specifically, the re-normalized level densities are derived through the formula

\begin{eqnarray}
\rho(U,J^{\pi})=\rho(U-\delta,J^{\pi})\exp(\alpha\sqrt{U-\delta}),
\label{eq0}
\end{eqnarray}
where the energy shift $\delta$ of the excitation energy $U$ is essentially extracted from the analysis of the cumulative number of experimental levels, and $\alpha$ from the experimental s-wave neutron spacings. Equation (\ref{eq0}) has been used to fit the 289 nuclides for which this experimental information exists. The corresponding $\delta$ and $\alpha$ values along with the experimental data are also given in BRUSLIB.
Figure \ref{nld} illustrates the calculated  $^{56}$Fe, $^{99}$Ru, $^{144}$Sm, and $^{209}$Bi NLDs for different spins and parities. For these same nuclides, Fig. \ref{level} shows that the predicted cumulative numbers of levels reproduce the measured values quite well.

\begin{figure}
\centering
\includegraphics[width=10.0cm]{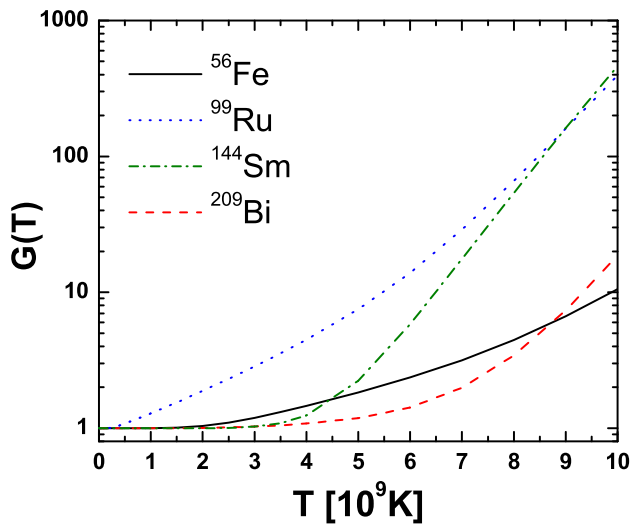}
\vskip-0.3cm
\caption{(Colour online) Partition functions for $^{56}$Fe, $^{99}$Ru, $^{144}$Sm, and $^{209}$Bi as a function of the temperature.}
\label{pf}
\end{figure}

The astrophysical reaction rates require the knowledge of the temperature-dependent partition function for each nuclide ($Z,A$). Its value normalized to the ground-state value is given by

\begin{eqnarray}
G(T)&=&\sum_{i}\frac{2J_{i}+1}{2J_{0}+1}\exp(-\frac{U_{i}}{kT}) \nonumber\\
&&+\int_{U_{th}}\sum_{J,\pi}\frac{2J+1}{2J_{0}+1}\rho(U,J,\pi)\exp(-\frac{U}{kT})dE,
\label{eq5}
\end{eqnarray}
where $J_{0}$ is the ground-state spin, and $U_{i}$ and $J_{i}$ are the excitation energy and spin for the i-$th$ discrete excited state. The summation extends over the experimentally known energy levels up to an excitation energy $U_{th}$, above which the knowledge of the energy spectrum is considered to be incomplete. The integration covers the energy range above $U_{th}$, and involves levels with all possible spins $J$ of parities $\pi = +$ and $-$. In practice the BRUSLIB partition functions are calculated using the experimental excited spectrum whenever available \citep{RIPL3} and, otherwise, the NLDs predicted by \citet{Goriely08}. BRUSLIB provides the partition functions for the same set of 8508 nuclides, and for temperatures in the 0.001 $\leq$ $T_{9}$ $\leq$ 10 range (where $T_{9}$ denotes the temperature in $10^{9}$K). For illustration, Fig. \ref{pf} shows the partition functions for $^{56}$Fe, $^{99}$Ru, $^{144}$Sm, and $^{209}$Bi.

\subsection{E1 strength function}

Gamma-ray strength functions are involved in the HF calculation of capture cross sections, $\gamma$-ray production spectra, isomeric state populations, and competition between $\gamma$-ray and particle emission. In this context, relevant multi-polarities are E1, M1, and E2 strengths, among which E1 dominates in general.

Large-scale derivations of the E1 strength function \citep{Goriely04} have been conducted with the use of the HFB plus  quasi-particle random phase approximation (QRPA) models \citep{Khan01} based on a realistic Skyrme interaction. The HFB model allows one to treat pairing effects on the ground state in a self-consistent way, while QRPA views the collective nuclear excitation as a collective superposition of two quasi-particle states built on top of the HFB ground state. This collective aspect of the excitation makes QRPA an accurate tool for investigating the E1-strength function in both closed and open shell nuclei \citep{Goriely04}. BRUSLIB contains the E1 strength functions obtained from the HFB plus QRPA models based on the BSk14 Skyrme force for all 8508 nuclides with $8\leq Z \leq 110$ lying between the two drip lines. Data are presented on an energy grid of 0.1 MeV between 0 and 30 MeV.

\begin{figure}
\centering
\includegraphics[width=10.0cm]{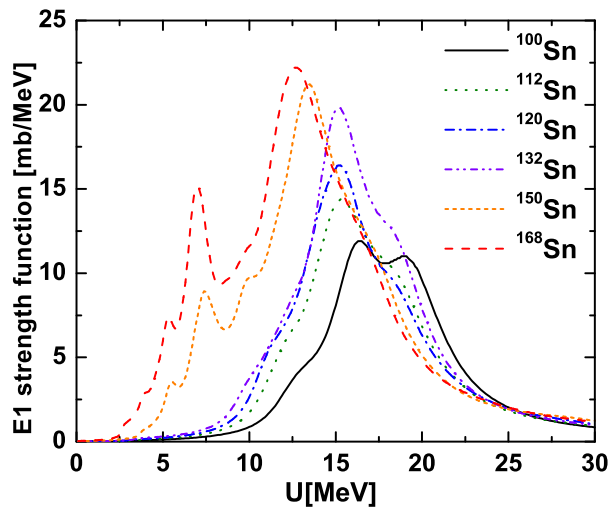}
\vskip-0.5cm
\caption{(Colour online) E1-strength functions of some Sn isotopes.}
\label{E1}
\end{figure}

Figure \ref{E1} illustrates the E1-strength functions for some Sn isotopes. In the neutron-deficient region as well as along the $\beta$-valley of stability, the strengths are very similar to the empirical Lorentzian-like approximation. In the neutron-rich region, the HFB+QRPA predictions start deviating from a simple Lorentzian shape with some extra strength located at an energy lower than the giant dipole resonance energy. The more exotic the nucleus, the stronger this low-energy component referred to as the "E1 pygmy resonance". This prediction agrees with existing experimental results  \citep{Goriely04}.

\subsection{Fission properties}

\begin{figure}
\centering
\vspace{-2.5cm}
\includegraphics[width=9.0cm]{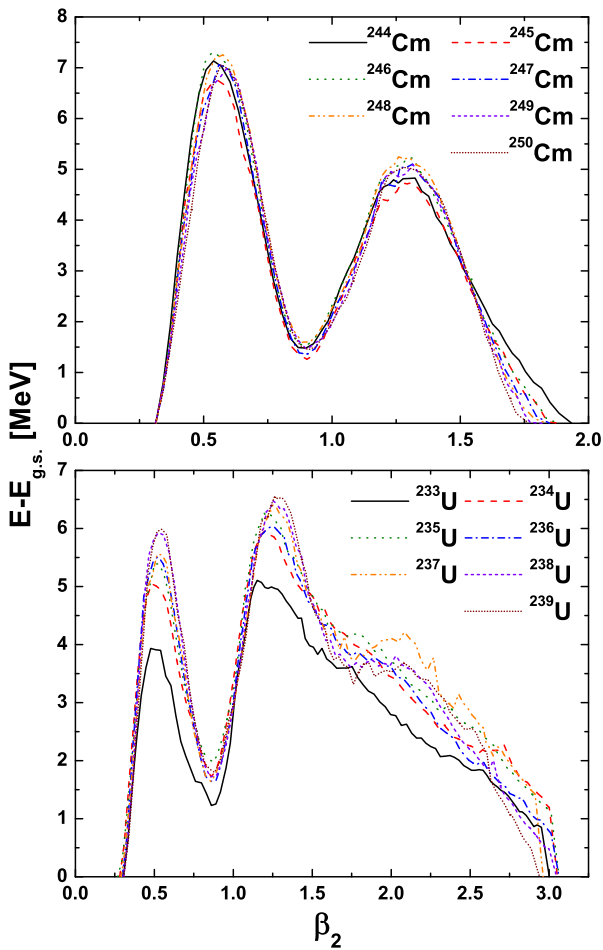}
\caption{(Colour online) Predicted HFB fission paths for the Cm and U isotopes close to the valley of $\beta$ stability versus the deformation parameter $\beta_2$.}
\label{fissionbr}
\end{figure}

Fission is of special importance in the modelling of the r-process of nucleosynthesis, which is invoked to explain the origin of approximately half of the stable nuclides above iron observed in nature, as well as the production of the long-lived actinides $^{232}$Th, $^{235}$U, and $^{238}$U, which are often used as cosmochronometers to estimate an upper limit of the age of the Galaxy. For this purpose, the probabilities of spontaneous, $\beta$-delayed and neutron-induced fission for the nuclides with $80 \leq Z \leq 110$ mainly located in the neutron-rich region of the nuclear chart are needed.

All key ingredients entering the fission description, such as the static fission path (including the height and width of the primary, secondary and possible tertiary barriers) and the NLD, are derived within the same microscopic model. The HFB model with the BSk14 Skyrme force is used to estimate the static fission path (allowing for axially symmetric deformations as well as left-right asymmetries) for all nuclides with $90\leq Z \leq 110$ lying between the valley of $\beta$-stability and the neutron drip lines \citep{Goriely09}. These predictions have been shown to agree with existing experimental data  \citep{Goriely09}.

Figure \ref{fissionbr} displays the fission paths for the U and Cm isotopes close to the valley of $\beta$ stability. Each fission path corresponds to the most gently climbing or steepest descending path found and projected along one deformation parameter, namely the quadrupole deformation $\beta_{2}$. Although a tiny third barrier clearly appears at strong deformations for some U isotopes, the fission path for these nuclei appears to be well-represented by a traditional double-humped barrier, at least in a local region located close to the saddle-point deformations. The situation can be quite different farther away from stability \citep{Goriely08c,RIPL3}. In fact, the fission path for exotic neutron-rich nuclei cannot be simply approximated by a double-humped barrier with parabolic shapes.

\begin{figure}
\centering
\vspace{-2.5cm}
\includegraphics[width=9.0cm]{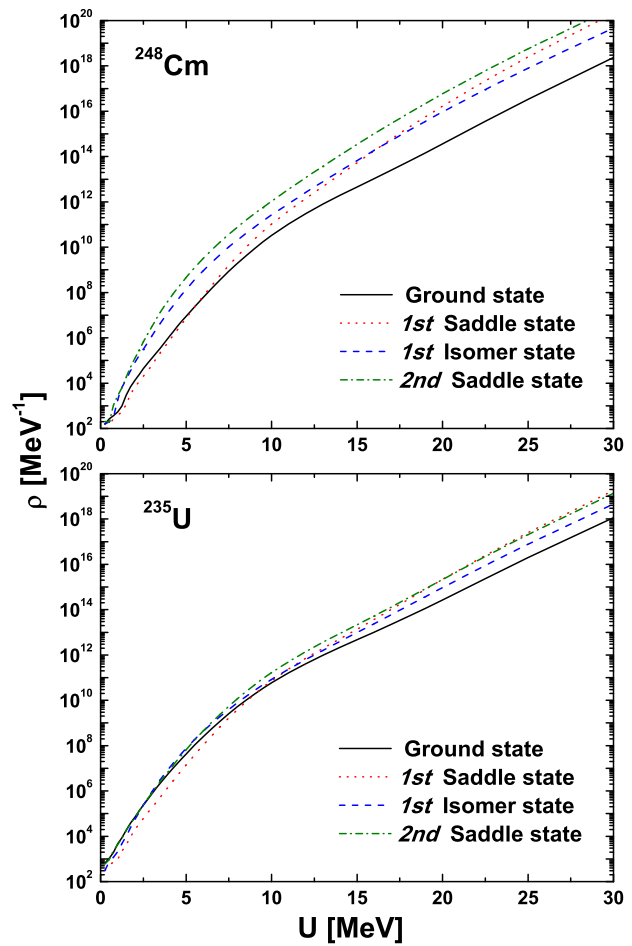}
\caption{(Colour online) Total NLDs for $^{248}$Cm and $^{235}$U at different locations along their fission path.}
\label{fissionlvlden}
\end{figure}

The HFB plus combinatorial method developed to estimate the NLD at ground-state deformation (Sect. 2.2) is also used to calculate the NLD at the saddle points along the fission path, making a coherent use of the corresponding HFB predictions for the single-particle level scheme and pairing strength at the corresponding deformation \citep{Goriely09}. Because of the lack of experimental information, the same prescription is used for the saddle points as for the ground state, i.e., a total of three phonons are coupled to the excitation configurations of a maximum of four particle-holes. Quadrupole, octupole, and hexadecapole phonons are included, their energies being assumed identical to those of the ground state. BRUSLIB provides the corresponding NLDs for $90 \leq Z \leq 110$ nuclides at each of their two (or three) highest saddle point barriers and one (or two) shape isomers, and at energies up to $U = 200~{\rm MeV}$ and spins up to $J = 49$ (99/2). As an example, the $^{248}$Cm and $^{235}$U total NLDs are shown in Fig. \ref{fissionlvlden} at different locations along the fission paths.

\section{HF reaction rates}

\begin{figure*}
\centering
\includegraphics[width=\textwidth]{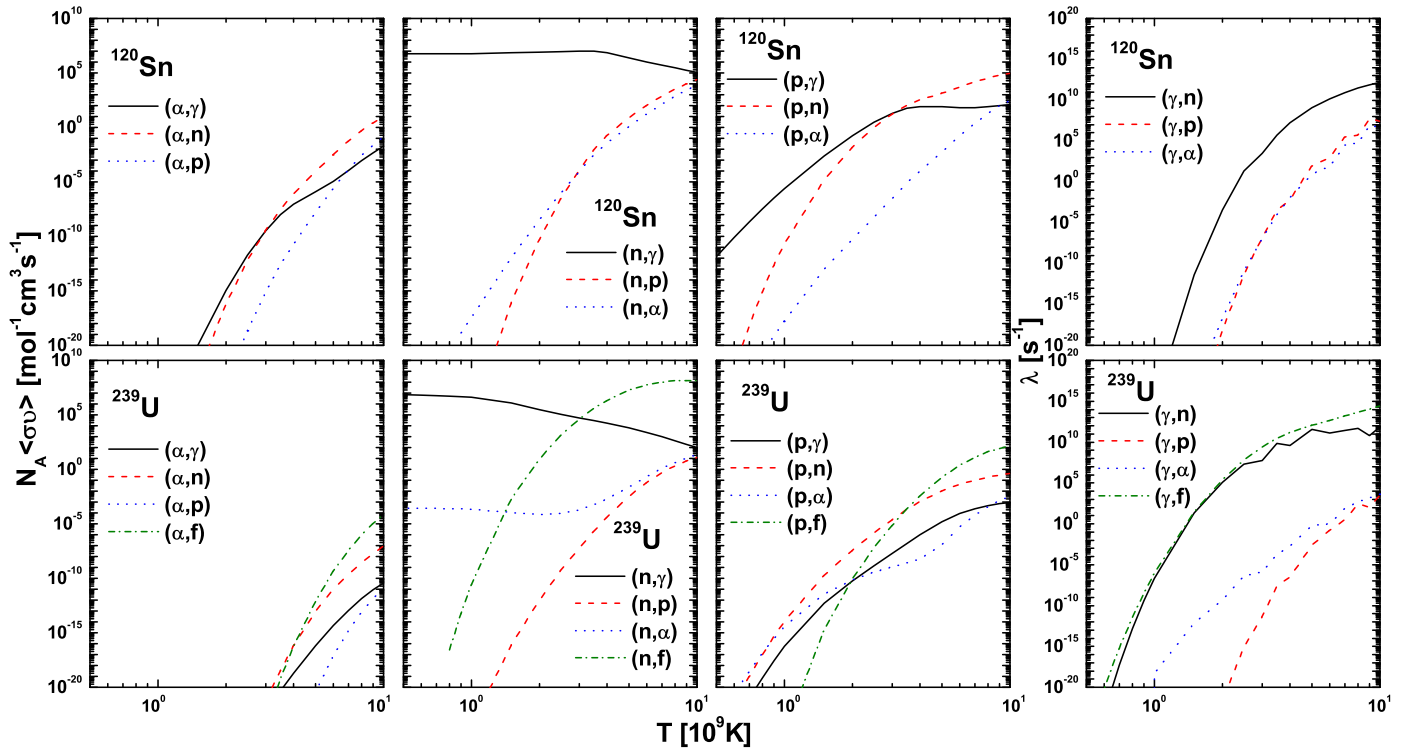}
\caption{BRUSLIB reaction rates on the targets $^{120}$Sn and $^{239}$U.}
\label{rate}
\end{figure*}

Thousands of reactions of astrophysics relevance involve excited states of the target, and more or less exotic nuclides. No wonder then that experimental data are largely missing, and will remain so for a long time to come. Theory has therefore to provide the necessary data.

The reaction rates supplied by BRUSLIB \citep{AA08} are obtained with the code TALYS, which is dedicated to nuclear reaction simulations through the use of many state-of-the-art nuclear models \citep{Koning02,Koning04,TALYS}. It is able to cover all main mechanisms at play in light particle-induced nuclear reactions, and provides a complete description of all reaction channels. In particular, TALYS takes into account most types of direct, pre-equilibrium, and compound mechanisms to estimate the total reaction probability, as well as the competition between the various open channels. The code is optimized for incident projectile energies ranging from 1 keV to 200~MeV on target nuclei with mass numbers between 12 and 339. It includes photon, neutron, proton, deuteron, triton, $^{3}$He, and $\alpha$-particles as both projectiles and ejectiles, and single-particle as well as multi-particle emissions and fission. All experimental information on nuclear masses, deformations, and low-lying states spectra from, e.g., the RIPL-3 database \citep{RIPL3} is adopted as input. If the necessary laboratory information is missing, the BRUSLIB data described in Sects. 2.1-2.4 are used. This concerns the nuclear masses, spins, parities, NLDs at ground state or fission saddle points, and partition functions. In addition, the E1 strengths (Sect. 2.3) were used to calculate the photon transmission coefficients $T_{\gamma}$, and the fission paths of Sect. 2.4 were used in conjunction with the WKB method \citep{WKB,Goriely09} to calculate the fission transmission coefficients $T_{f}$. As for the particle emission channels, the transmission coefficients were evaluated on the basis of the optical potential of \citet{koning03} for the nucleon-nucleus interaction and of \citet{Demetriou02} for the $\alpha$-nucleus interaction.

BRUSLIB supplies about 100000 TALYS rates for (n,$\gamma$), (n,p), (n,$\alpha$), (p,$\gamma$), (p,n), (p,$\alpha$), ($\alpha$,$\gamma$), ($\alpha$,n), ($\alpha$,p), ($\gamma$,n), ($\gamma$,p), and ($\gamma$,$\alpha$) reactions on some 8500 targets with $8\leq Z \leq 110$ lying between the proton and the neutron drip lines, and for temperatures in the 0.001 $\leq$ $T_{9}$ $\leq$ 10 range. For nuclides with $80 \leq Z \leq 110$ that lie between the proton and the neutron drip lines, each of the neutron-, proton-, $\alpha$-, and $\gamma$-induced fission rates, i.e. the channels (n,f), (p,f), ($\alpha$,f), and ($\gamma$,f), can be accessed from BRUSLIB. For $Z\ge90$, they are calculated on the basis of the HF model making use of the fission barriers and the ground-state and saddle point NLDs calculated as described in Sect. 2.4. As an example, Fig. \ref{rate} shows the various reaction rates on the targets $^{120}$Sn and $^{239}$U.

The TALYS estimates for n-, p-, deuton-, triton-, $^{3}$He-, and $\alpha$-induced reaction rates on targets from Li to Na that could be involved in Big-Bang nucleosynthesis \citep{BBN} are also included in BRUSLIB. These predictions can be viewed at best as rough first approximations when no information from other sources is available. As summarized by \citet{BBN}, TALYS is expected to provide these reaction rate predictions within a 3 orders of magnitude accuracy in the temperature range of Big-Bang nucleosynthesis, though in many cases the agreement can be regarded as satisfactory.

\begin{figure}
\centering
\includegraphics[width=10.0cm]{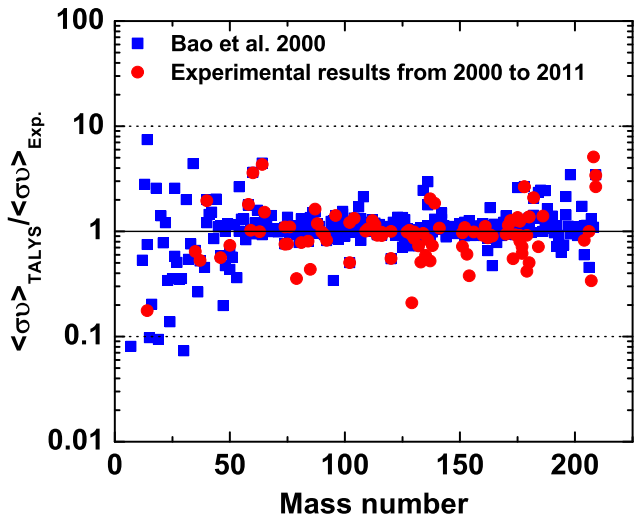}
\caption{(Colour online) Comparison between the $T_{9}$ = 0.3 BRUSLIB radiative neutron capture rates and the experimental rates provided by the compilation of \citet{Bao00}, complemented with data published in refereed journals between 2000 to 2011 (references mentioned in Sect. 5).}
\label{ratecompare}
\end{figure}

More generally, the uncertainties in the calculated cross sections arise from two main origins. The first one is related to the reaction mechanism itself. All codes previously developed for astrophysical applications only describe one reaction mechanism, either of direct or of compound nucleus type, and include a number of approximations. Even if many of these can be justified by the very low relative energies (often a few keV to 1 MeV) of astrophysical relevance, it is of interest to test their validity on a quantitative basis. TALYS is suited for this, in particular through the inclusion of \citep{AA08}
\begin{itemize}
\item the pre-equilibrium reaction mechanism;
\item the detailed description of the decay scheme, including the description of
$\gamma$-delayed particle emission and the possible particle emission from all
residual nuclei;
\item the multi-particle emission;
\item the choice of a variety of width-fluctuation correction factors;
\item the coupled channel description for deformed nuclei;
\item the fission channel for the compound as well as for the residual nuclei.
\end{itemize}
Other uncertainties come from the nuclear ingredients that enter the reaction model. When dealing with nuclear astrophysics applications, these ingredients should be derived from global, universal, and microscopic models. The large number of nuclei involved in modelling of some nucleosynthesis mechanisms indeed implies that global models should be used. On the other hand, a universal description of all nuclear properties within a unique framework for all nuclides involved in a nuclear network ensures the essential coherence of the predictions for all unknown data. Finally, a microscopic description provided by a physically sound theory based on first principles likely renders extrapolations away from experimentally known energy or mass regions more reliable than predictions derived from more or less parametric approaches of various types and levels of sophistication. The microscopic models used in TALYS to calculate the ingredients of the BRUSLIB reaction rates meet these criteria through a satisfactory compromise between accuracy and reliability. This would make the BRUSLIB reaction rate library unique away from stability, while it is just of comparable reliability close to stability, as illustrated by Fig. \ref{ratecompare}. The full BRUSLIB database can be found at {\it http://www.astro.ulb.ac.be/bruslib}

\section{Experiment-based reaction rates: the NACRE II compilation}

Since the NACRE compilation \citep{NACRE}, many charged-particle-induced cross sections of astrophysical interest have been measured or re-measured. In addition, it has been considered desirable to invest more effort than into NACRE on the modelling of the extrapolations of the experimental data to the very low energies of astrophysical relevance. These extrapolations are essential because the relevant energy domain cannot be reached in the laboratory, except in very few cases. This situation has motivated the launch of the NACRE II compilation, which aims at revising some of the NACRE rates \citep{nacre2pro1,nacre2pro2,nacre2pro3,nacre2pro4,NACRE2}.

In its current version, NACRE II surveys 34 two-body exothermic reactions (15 particle-transfer and 19 radiative capture reactions) on targets with $A \leq 16$ listed in Table \ref{nacre2}. Phenomenological potential models are adopted to describe and extrapolate their resonant and non-resonant cross sections at very low energies. The HF approximation is not suited for the cases under consideration, except possibly at the highest considered energies (or temperatures).

\begin{table}
\caption{Reactions included in NACRE II.}
\label{nacre2}
\centering
\begin{tabular}{cccc}
\hline\hline
No. & reaction &  No. & reaction\\
\hline
1    &  $^{2}$H(p,$\gamma$)$^{3}$He  &           18   &
$^{9}$Be(p,$\gamma$)$^{10}$B     \\
2    &  $^{2}$H(d,$\gamma$)$^{4}$He  &           19   &
$^{9}$Be(p,d)2$^{4}$He           \\
3    &  $^{2}$H(d,n)$^{3}$He         &           20   &
$^{9}$Be(p,$\alpha$)$^{6}$Li     \\
4    &  $^{2}$H(d,p)$^{3}$H          &           21   &
$^{9}$Be($\alpha$,n)$^{12}$C     \\
5    &  $^{2}$H($\alpha$,$\gamma$)$^{6}$Li    &  22   &
$^{10}$B(p,$\gamma$)$^{11}$C     \\
6    &  $^{3}$H(d,n)$^{4}$He         &           23   &
$^{10}$B(p,$\alpha$)$^{7}$Be     \\
7    &  $^{3}$H($\alpha$,$\gamma$)$^{7}$Li  &    24   &
$^{11}$B(p,$\gamma$)$^{12}$C     \\
8    &  $^{3}$He(d,p)$^{4}$He  &                 25   &
$^{11}$B(p,$\alpha$)2$^{4}$He     \\
9    &  $^{3}$He($^{3}$He,2p)$^{4}$He  &         26   &
$^{11}$B($\alpha$,n)$^{14}$N     \\
10   &  $^{3}$He($\alpha$,$\gamma$)$^{7}$Be  &   27   &
$^{12}$C(p,$\gamma$)$^{13}$N     \\
11   &  $^{6}$Li(p,$\gamma$)$^{7}$Be  &          28   &
$^{12}$C($\alpha$,$\gamma$)$^{16}$O    \\
12   &  $^{6}$Li(p,$\alpha$)$^{3}$He  &          29   &
$^{13}$C(p,$\gamma$)$^{14}$N     \\
13   &  $^{7}$Li(p,$\gamma$)$^{8}Be$  &          30   &
$^{13}$C($\alpha$,n)$^{16}$O     \\
14   &  $^{7}$Li(p,$\alpha$)$^{4}$He  &          31   &
$^{13}$N(p,$\gamma$)$^{14}$O     \\
15   &  $^{7}$Li($\alpha$,$\gamma$)$^{11}$B  &   32   &
$^{14}$N(p,$\gamma$)$^{15}$O     \\
16   &  $^{7}$Be(p,$\gamma$)$^{8}$B &            33   &
$^{15}$N(p,$\gamma$)$^{16}$O     \\
17   &  $^{7}$Be($\alpha$,$\gamma$)$^{11}$C &    34   &
$^{15}$N(p,$\alpha$)$^{12}$C     \\
\hline\hline
\end{tabular}
\end{table}

The methodology adopted for the construction of NACRE II can be summarized as follows:

\noindent\textit{Reaction mechanisms} Two main mechanisms are considered for low-energy nuclear reactions: the compound nucleus and the direct reaction processes. The direct reaction process is important, and often dominant, in charged-particle-induced reactions at the very low energies of astrophysical interest. Given the difficulty to tunnel through the Coulomb barrier, the reaction may occur before the projectile can penetrate deep inside the target nucleus. The formation of a compound state is accordingly suppressed, which is reinforced especially in light nuclei by the paucity of quasi-bound levels.

\noindent\textit{Adopted models} The potential model, including the E1, E2, and M1 transitions, is used to evaluate the 19 capture reactions, while the DWBA with zero-range interaction approximation is employed to study the 15 transfer reactions. The adopted nuclear potential is a real Woods-Saxon potential with a surface absorption imaginary part. At low energies, the significant direct and various discrete resonant contributions to the cross sections are both taken into account in these two models. However, for a few cases involving extremely narrow resonances, the Breit-Wigner formula is adopted for evaluating the corresponding reaction rates around the resonance energy.

\noindent\textit{Selection of experimental data} The primary ensemble of experimental low-energy cross section data of NACRE II comprises those included in NACRE and supplementary ones that have been published by the end of 2011 preferentially in refereed journals. Generally speaking, we take the selected experimental data on cross sections and associated errors at face value since we are not confident enough to do otherwise, given the quite limited information that is normally available to us. We also do not refer in NACRE II to any differential quantities, even when they have been measured.

\noindent\textit{Fitting procedure} For transfer reactions, a DWBA model is adopted with a cut-off energy for the free parameter fit on the high-energy side normally set at $E_{c.m.}$ = 1 MeV. However, when resonances are found experimentally below 1 MeV, only the one with the lowest excitation energy is embraced into the fit. This procedure is justified because the S-factor below that energy range normally behaves like the tail of that resonance. Just for the practical purpose of reproducing the measured cross section data, one may drastically reduce the number of potential parameters of the entrance and exit channels and of the form factor without causing much damage. First of all, for the entrance channel, we generally retain only a shallow, imaginary part of the nuclear potential, which takes into account the weak absorption to the exit channel by particle transfer. We then parameterize rather than fit the radii and diffuseness of the global potentials. The potential depths are left as the adjustable parameters to be fitted. We also treat the spectroscopic factor as an adjustable parameter for the absolute value of the cross section. In general, the best values of the adjustable parameters have been derived by applying the standard $\chi^{2}$ fit with the occasional help of a fit-by-eye.

A potential model is used for the calculation of radiative capture cross sections. These are rarely measured below $E_{c.m.}$ = 0.1 MeV, where  non-resonant contributions may well be dominant. Therefore we adopt a strategy of parameterization that is quite different from the one selected for transfer reactions. In particular, we try to fit the resonances, if any, with the real potential form to deduce the non-resonant contributions simultaneously. To reduce the number of parameters, we take the same set of radii and diffuseness for both the initial and final states. Hence, for a given resonant or non-resonant contribution, the potential depth, radii, diffuseness, and the re-normalization constant including the spectroscopic factor are left to be fitted. This is quite in contrast to the strategy used for transfer reactions, and fit-by-eye is used more often than in the case of transfer reactions.

\noindent\textit{Reaction rate evaluation} For each reaction, three different sets of model parameters are determined, duly taking into account reported experimental uncertainties. From these sets, the adopted, low, and high limits of the extrapolated S-factors are calculated by the reaction model. Reaction rates $N_{A}<\sigma\upsilon>$ are evaluated in the $0.001 \leq T_{9} \leq 10$ temperature range. All details can be found at {\it http://www.astro.ulb.ac.be/nacreii}.

Many more details on NACRE II than those provided above can be found in \citet{NACRE2}. In particular, specific comments are given for each of the reactions of Table \ref{nacre2} along with the values of all model parameters used to calculate the tabulated adopted, lower and upper limits of the rates. The NACRE II rates are also compared to the NACRE rates.

\section{NETGEN}

\subsection{Content}

The nuclear NETwork GENerator NETGEN v10.0 is an interactive, web-based tool to help astrophysicists in building up a nuclear reaction network as defined by each user. It generates tables of the necessary nuclear reaction rates on a temperature grid specified by the user, and provides the references of the sources of these rates. All reaction rates, theoretical or experiment-based, include the contribution of thermally populated excited target states through the calculation of the stellar enhancement factor.

The following nuclear data are available in NETGEN for about 8600 nuclides located between the neutron and proton drip lines in the $1\leq Z \leq 110$ range:
\vskip-0.4truecm
\begin{itemize}
\item for compiled experiment-based nuclear reaction rates:
\vskip0.2truecm
   \begin{itemize}
   \item the NACRE II data (see Sect. 4) for $1 \leq A \leq 16$ targets;
   \item the NACRE data  \citep{NACRE}, when not superseded by NACRE II, or by other more recent compilations than NACRE listed below;
   \item the compilation of the rates of Big-Bang reactions based on the R-matrix model \citep{Dec04};
   \item the 2010 evaluation of Monte Carlo-based rates of resonant proton capture reactions on targets in the
 20 $\leq$ A $\leq$ 40 range \citep{Iliadis10};
   \item the proton-induced reaction rates on stable and unstable target nuclei in the 20 $<$ A $<$ 40 region \citep{Iliadis01};
   \item when not available in the above-mentioned compilations, the reaction rates proposed by \citet{CF88};
   \item the compilation of \citet{Bao00} for radiative neutron capture rates;
   \item various radiative neutron capture rates made available between 2000 and 2011 that supersede the  compilation of \citet{Bao00}
\citep{ncap51,ncap29,ncap12,ncap14,ncap36,ncap25,ncap45,ncap47,ncap30,ncap32,
ncap34,ncap37,ncap40,ncap15,ncap53,
ncap23,ncap02,ncap39,ncap42,ncap04,ncap06,ncap11,ncap13,ncap57,ncap27,ncap44,
ncap46,ncap55,ncap48,ncap07,ncap56,ncap03,
ncap31,ncap19,ncap20,ncap16,ncap22,ncap18,ncap38,ncap09,ncap05,ncap41,ncap43,
ncap49,ncap54,ncap35,ncap52,ncap33,
ncap08,ncap01,ncap10,ncap50,ncap17,ncap21,ncap24,ncap26,ncap28};
    \item additional specific experimental reaction rates published in refereed journals that were not included in any of the above compilations.
    \end{itemize}
\end{itemize}

\begin{itemize}
\item for theory based nuclear reaction rates:
the rates of the neutron-, proton-, $\alpha$-, and $\gamma$-induced reaction rates, including fission, provided by BRUSLIB (Sect. 2).
\end{itemize}

\begin{itemize}
\item for compiled experimental and theoretical $\beta$-decay and electron capture rates:
\vskip0.2truecm
  \begin{itemize}
  \item the experimental $\beta$-decay rates of \citet{Katakura96};
  \item the $\beta$-decay rates of \citet{Takahashi87} \citep[see also][]{Goriely99};
  \item specific electron capture rates from various published sources;
  \item $\beta$-decay rates calculated with the revised Gross theory \citep{Tachibana90} and Q-values from the HFB-14 mass model;
  \item $\beta$-decay rates calculated with the QRPA plus FRDM models by \citet{Moller96};
  \item the density- and temperature-dependent $\beta$-decay rates calculated by \citet{Oda94} or \citet{Langanke00}.
  \end{itemize}
\end{itemize}

\subsection{Instructions for use of NETGEN}

For each nuclear reaction, $\beta$-decay and electron-capture rate on a nuclide involved in a network defined by the user, NETGEN adopts by default the data source that is considered to be the most reliable. In order of preference, it selects the latest available compilation based on experimental data, followed by the BRUSLIB rates. The user may nevertheless make another choice for selected cases by specifying a \textit{bibliographic index} for each reaction. The possible choices for this bibliographic index are given in a help panel on the website and in the output log file. NaN (Not a Number) flags are present in the output if the temperature grid requested by the user extends beyond the limit of validity of the experimental range. To remove these NaN, NETGEN provides the option to use theory-based reaction rates properly scaled (to ensure continuity). Finally, a FORTRAN computer program handling the rates is also provided. More details can be found at {\it http://www.astro.ulb.ac.be/Netgen}.

The general steps for constructing a network and the associated reaction rate library are as follows:
\vskip0.2truecm
\noindent\textit{Step 1: Network preparation} A network may be generated by four non-mutually exclusive methods:
\begin{itemize}
\item reactions typed in one by one. When choosing a reaction, the bibliographic index should also be given;
\item automatically generated network in the ($N,Z$) plane. The nuclides involved can be selected by charge ($Z$) and mass ($A$) number, boundaries (drip lines, stability line), and among various possible sets (proton-, neutron-, $\alpha$-captures, photo-dissociations, etc.). In this option, the default reaction rates (\textit{bibliographic index} = 1) are selected. This means that if several rates are available for a given reaction, the experimental or experiment-based rate is chosen rather than the theoretical one. If several experimental rates are provided, the most recently published one is selected. For $\beta$-decays, the rates are taken from \citet{Takahashi87}. If a rate is not available from this source, other experimental and most recent rates are chosen;
\item upload the reaction file according to the NETGEN prescription;
\item select the reactions from the specific pre-defined database.
\end{itemize}

\noindent\textit{Step 2: Network construction} After step 1, the user has the possibility to list the current network, modify it, sort it, or clear it. Plotting the network in the ($N,Z$) plane and individual reaction rates is offered as additional options.

\noindent\textit{Step 3: Compute the reaction rates} The network reaction rates are computed on the user-defined temperature grid. A file recording the rates is generated, as well as a log file that provides the references from which the rates have been calculated. Plots of the rates versus temperature can be obtained.

\section{Summary and outlook}

The last update of the nuclear databases BRUSLIB and NACRE II, and of the nuclear network generator NETGEN was presented. This package is of unprecedented scope and quality. It is accessible through the website {\it http://www.astro.ulb.ac.be/NuclearData}.

BRUSLIB contains an immense variety of static properties for nuclei that lie between the proton and neutron drip lines over the range $8\leq Z \leq 110$, as well as the rates of proton-, neutron-, and $\alpha$-particle-induced reactions on these nulides. Calculated rates of interest for Big-Bang nucleosynthesis are also given, but have to be considered as rough guesses only. BRUSLIB has the unique feature that this vast body of data are predicted by a global and universal microscopic model, and by the modern reaction code TALYS. All BRUSLIB data can be accessed via user-interactive figures and tables.

NACRE II supersedes NACRE for 15 transfer and 19 radiative capture reactions on nuclides with $A < 16$. It includes experimental data published after the release of NACRE and up to 2011, and makes use of a phenomenological model to extrapolate the experimental data to the very low energies of astrophysical relevance, and to evaluate the uncertainties on the rates. Detailed electronic material is supplied, in particular, tabulated rates.

The web-based nuclear network generator NETGEN gives the possibility to set up a network, and to obtain the necessary network (nuclear and weak) reaction rates in a tabular form on a grid selected by the user, and from sources that are selected by default, or by the user.

The databases described in this work of course need continuous updating and improvement. Just as an illustration, TALYS offers the possibility of using alternative local or global models for predicting various nuclear ingredients. Additional effort should be put in the study of not only parameter uncertainties, but also of model uncertainties (e.g. the $\alpha$-particle optical potential) that affect the astrophysical reaction rates \citep{Arnould06,Arnould07}. In addition, direct captures not treated in TALYS should be scrutinized on a large-scale basis with the help of a global model using of the same ingredients as those used in HF calculations. The release of an extended spallation reaction library is also envisioned.

\begin{acknowledgements}
We warmly thank K. Takahashi for his key contribution to the NACRE II compilation. This work has been supported by the Communaut\'e Fran\c caise de Belgique
(Actions de Recherche Concert\'ees). Y.X. is post-doctoral research fellow from F.R.S.-FNRS.
SG. is F.R.S.-FNRS Research Associate.
\end{acknowledgements}


\begin{thebibliography}{116}

\bibitem [Abbondanno et al., 2004] {ncap51} Abbondanno, U., Aerts, G.,
Alvarez-Velarde, F., et al. (n-TOF Collaboration) 2004, Phys. Rev. Lett., 93,
161103
\bibitem [Aerts et al., 2006] {ncap29} Aerts, G., Abbondanno, U., Alvarez, H.,
et al. (n-TOF Collaboration) 2006, Phys. Rev. C, 73, 054610
\bibitem [Aikawa et al., 2005] {AA05}  Aikawa, M., Arnould, M., Goriely, S.,
Jorissen, A., \& Takahashi, K. 2005, A\&A, 441, 1195
\bibitem [Aikawa et al., 2006] {nacre2pro1} Aikawa, M., Arai, K., Arnould, M., Takahashi, K., Utsunomiya, H., in: Harissopulos, S. V., Demetriou, P., Julin, R. (Eds.), Proceedings of International Conference  on Frontiers in Nuclear Structure, Astrophysics and Reactions (FINUSTAR), AIP Conf. Proc., 831 (2006), 26-30
\bibitem [Angulo et al., 1999] {NACRE} Angulo, C., Arnould, M., Rayet, M.,
Descouvemont, P., Baye, D., Leclercq-Willain, C., Coc, A., Barhoumi, S., Auger,
P., Rolfs, C., Kunz, R., Hammer, J. W., Mayer, A., Paradellis, T., Kossionides,
S., Chronidou, C., Spyrou, K., Degl'Innocenti, S., Fiorentini, G., Ricci, B.,
Zavatarelli, S., Providencia, C., Wolters, H., Soares, J., Grama, C., Rahighi,
J., Shotter, A., \& Lamehi-Rachti, M. 1999, Nucl. Phys. A, 656, 3 (NACRE)
\bibitem [Arnould \& Goriely, 2006] {Arnould06} Arnould, M., \& Goriely, S.
2006, Nucl. Phys. A, 777, 157c
\bibitem [Arnould \& Katsuma, 2007] {nacre2pro3} Arnould, M., \& Katsuma, M., in: Bersillon, O., Gunsing, F., Bauge, E., Jacqmin, R. J., Leray, S. (Eds.), Proceedings of International Conference  on Nuclear Data for Science and Technology (ND2007), Nice, France, April 22-27, 2007, EDP Sciences (Paris, 2008), 5-10
\bibitem [Arnould et al., 2007] {Arnould07} Arnould, M., Goriely, S., \&
Takahashi, K. 2007, Phys. Rep., 450, 97
\bibitem [Audi, 2010] {Audi10} Audi, G. 2010, Atomic Mass Evaluation 2010,
Private communication
\bibitem [Bao et al., 2000] {Bao00} Bao, Z. Y., Beer, H., K\"{a}peler, F., Voss,
F., Wisshak, K., \& Rauscher, T. 2000, Atomic Data Nucl. Data Tables, 75, 1
\bibitem [Bauge et al., 2001] {Bauge01} Bauge, E., Delaroche, J. P., \& Girod,
M. 2001, Phys. Rev. C, 63, 024607
\bibitem [Best et al., 2001] {ncap12} Best, J., Stoll, H., Arlandini, C., et al.
2001, Phys. Rev. C, 64, 015801
\bibitem [Blackmon et al., 2002] {ncap14} Blackmon, J. C., Raman, S., Dickens,
J. K., et al. 2002, Phys. Rev. C, 65, 045801
\bibitem [Borella et al., 2007] {ncap36} Borella, A., Gunsing, F., Moxon, M.,
Schillebeeckx, P., Siegler, P. 2007, Phys. Rev. C, 76, 014605
\bibitem [Capote et al., 2009] {RIPL3} Capote, R., Herman, M., Oblozinsky, P.,
Young, P. G., Goriely, S., Belgya, T., Ignatyuk, A. V., Koning, A. J., Hilaire,
S., Plujko, V. A., Avrigeanu, M., Bersillon, O., Chadwick, M. B., Fukahori, T.,
Ge, Z. G., Han, Y. L., Kailas, S., Kopecky, J., Maslov, V. M., Reffo, G., Sin,
M., Soukhovitskii, E. S., \& Talou, P. 2009, Nuclear Data Sheets, 110, 3107
\bibitem [Caughlan \& Fowler, 1988] {CF88} Caughlan, G. E., \& Fowler, W. A.
1988, Atomic Data Nucl. Data Tables, 40, 283
\bibitem [Chamel et al., 2008] {pro9} Chamel, N., Goriely, S., \& Pearson, J. M.
2008, Nucl. Phys. A, 812, 72
\bibitem [Coc et al., 2011] {BBN} Coc, A., Goriely, S.,
Xu, Y., Saimpert, M., \& Vangioni, E. 2012, ApJ, 744, 158
\bibitem [Cyburt et al., 2010] {REACLIB} Cyburt, R. H., Amthor, A. M., Ferguson, R., Meisel, Z., Smith, K., Warren, S., Heger, A., Hoffman, R. D., Rauscher, T., Sakharuk, A., Schatz, H., Thielemann, F. K., \& Wiescher, M. 2010, ApJS, 189, 240
\bibitem [Demetriou et al., 2002] {Demetriou02} Demetriou, P., Grama, C., \&
Goriely, S. 2002, Nucl. Phys. A, 707, 253
\bibitem [Descouvemont et al., 2004] {Dec04} Descouvemont, P., Adahchour, A.,
Angulo, C., Coc, A., \& Vangioni-Flam, E. 2004, At. Data Nucl. Data Tables, 88,
203
\bibitem [Dillmann et al., 2006a] {KADoNiS} Dillmann, I., Heil, M., K\"{a}ppeler,
F., Plag, R., Rauscher, T., \& Thielemann, F. K. 2006, AIP Conf. Proc., 819, 123, online at http://www.kadonis.org
\bibitem [Dillmann et al., 2006b] {ncap25} Dillmann, I., Heil, M., K\"{a}ppeler,
F., Rauscher, T., \& Thielemann, F. K. 2006, Phys. Rev. C, 73, 015803
\bibitem [Dillmann et al., 2009] {ncap45} Dillmann, I., Domingo-Pardo, C., Heil,
M., et al. 2009, Phys. Rev. C, 79, 065805
\bibitem [Dillmann et al., 2010] {ncap47} Dillmann, I., Domingo-Pardo, C., Heil,
M., et al. 2010, Phys. Rev. C, 81, 015801
\bibitem [Domingo-Pardo et al., 2006a] {ncap30} Domingo-Pardo, C., Abbondanno,
U., Aerts, G., et al. (n-TOF Collaboration) 2006a, Phys. Rev. C, 74, 025807
\bibitem [Domingo-Pardo et al., 2006b] {ncap32} Domingo-Pardo, C., Abbondanno,
U., Aerts, G., et al. (n-TOF Collaboration) 2006b, Phys. Rev. C, 74, 055802
\bibitem [Domingo-Pardo et al., 2007a] {ncap34} Domingo-Pardo, C., Abbondanno,
U., Aerts, G., et al. (n-TOF Collaboration) 2007a, Phys. Rev. C, 75, 015806
\bibitem [Domingo-Pardo et al., 2007b] {ncap37} Domingo-Pardo, C., Abbondanno,
U., Aerts, G., et al. (n-TOF Collaboration) 2007b, Phys. Rev. C, 76, 045805
\bibitem [Esch et al., 2008] {ncap40} Esch, E. I., Reifarth, R., Bond, E. M., et
al. 2008, Phys. Rev. C, 77, 034309
\bibitem [Fowler et al., 1967] {FCZ67} Fowler, W. A., Caughlan, G. E., \&
Zimmerman, B. A. 1967, Ann. Rev. Astr. Astro, 5, 525
\bibitem [Goriely, 1999] {Goriely99} Goriely, S. 1999, A\&A, 342, 881
\bibitem [Goriely et al., 2002]{pro3} Goriely, S., Samyn, M., Heenen, P. H.,
Pearson, J. M., \& Tondeur, F. 2002, Phys. Rev. C, 66, 024326
\bibitem [Goriely et al., 2003] {pro4} Goriely, S., Samyn, M., Bender, M., \&
Pearson, J. M. 2003, Phys. Rev. C, 68, 054325
\bibitem [Goriely et al., 2004] {Goriely04} Goriely, S., Khan, E., \& Samyn, M.
2004, Nucl. Phys. A, 739, 331
\bibitem [Goriely et al., 2005] {pro6} Goriely, S., Samyn, M., Pearson, J. M.,
\& Onsi, M. 2005, Nucl. Phys. A, 750, 425
\bibitem [Goriely et al., 2006] {pro7} Goriely, S., Samyn, M., \& Pearson, J. M.
2006, Nucl. Phys. A, 773, 279
\bibitem [Goriely et al., 2007] {pro8} Goriely, S., Samyn, M., \& Pearson, J. M.
2007, Phys. Rev. C, 75, 064312
\bibitem [Goriely et al., 2008a] {AA08} Goriely, S., Hilaire, S., \& Koning, A.
J. 2008a, A\&A, 487, 767
\bibitem [Goriely et al., 2008b] {Goriely08} Goriely, S., Hilaire, S., \&
Koning, A. J. 2008b, Phys. Rev. C, 78, 064307
\bibitem [Goriely et al., 2008c] {Goriely08c} Goriely, S., Pearson J.M.  2008c, in Proc. of Nuclear Data for Science \& Technology  (eds O. Bersillon, F. Gunsing, E. Bauge, R.Jacqmin, and S. Leray), EDP Sciences,  p. 203
\bibitem [Goriely et al., 2009a] {pro10} Goriely, S., Chamel, N., \& Pearson, J.
M. 2009a, Phys. Rev. Lett., 102, 152503
\bibitem [Goriely et al., 2009b] {Goriely09} Goriely, S., Hilaire, S., Koning,
A. J., Sin, M., \& Capote, R. 2009b, Phys. Rev. C, 79, 024612
\bibitem [Goriely et al., 2010] {Goriely10} Goriely, S., Chamel, N., \& Pearson,
J. M. 2010, Phys. Rev. C, 82, 035804
\bibitem [Guber et al., 2002] {ncap15} Guber, K. H., Sayer, R. O., Valentine, T.
E., et al. 2002, Phys. Rev. C, 65, 058801
\bibitem [Guber et al., 2010] {ncap53} Guber, K. H., Derrien, H., Leal, L. C.,
et al. 2010, Phys. Rev. C, 82, 057601
\bibitem [Heil et al., 2005] {ncap23} Heil, M., Dababneh, S., Juseviciute, A.,
et al. 2005, Phys. Rev. C, 71, 025803
\bibitem [Heil et al., 2008a] {ncap02} Heil, M., Winckler, N., Dababneh, S., et
al. 2008a, ApJ, 673, 434
\bibitem [Heil et al., 2008b] {ncap39} Heil, M., K\"{a}ppeler, F., Uberseder,
E., Gallino, R., \& Pignatari, M. 2008b, Phys. Rev. C, 77, 015808
\bibitem [Heil et al., 2008c] {ncap42} Heil, M., K\"{a}ppeler, F., Uberseder,
E., et al. 2008c, Phys. Rev. C, 78, 025802
\bibitem [Hilaire et al., 1998] {Hilaire98} Hilaire, S., Delaroche J. P., \&
Koning, A. J. 1998, Nucl. Phys. A, 632, 417
\bibitem [Hilaire et al., 2001] {Hilaire01} Hilaire, S., Delaroche J. P., \&
Girod, M. 2001, Euro. Phys. J. A, 12, 169
\bibitem [Hilaire \& Goriely, 2006] {Hilaire06} Hilaire, S., \& Goriely, S.
2006, Nucl. Phys. A, 779, 63
\bibitem [Iliadis et al., 2001] {Iliadis01} Iliadis, C., D'Auria, J. M.,
Starrfield, S.,	Thompson, W. J., \& Wiescher, M. 2001, ApJS, 134, 151
\bibitem [Iliadis et al., 2010] {Iliadis10} Iliadis, C., Longland, R.,
Champagne, A. E., Coc, A., \& Fitzgerald, R. 2010, Nucl. Phys. A, 841, 31
\bibitem [Jeukenne et al., 1977] {Jeukenne77} Jeukenne, J. P., Lejeune, A., \&
Mahaux, C. 1977, Phys. Rev. C, 16, 80
\bibitem [Katakura, 1996] {Katakura96} Katakura, J. 1996, Chart of the Nuclides,
Japanese Nuclear Data Committee and Nuclear Data Center, Japan Atomic Energy
Research Institute
\bibitem [Katoh et al., 2003] {ncap04} Katoh, T., Nakamura, S., Furutaka, K., et
al. 2003, J. of Nucl. Science \& Technology, 40, 559
\bibitem [Katsuma, 2006] {nacre2pro2} Katsuma, M., in: Arnould, M., Lewitowicz, M., Emling, H., Akimune, H., Ohta, M., Utsunomiya, H., Wada, T., Yamagata, T. (Eds.), Proceedings of Tours Symposium on Nuclear Physics VI, Tours, France, September 5-8, 2006, AIP Conf, Proc., 891, 355-363
\bibitem [Khan et al., 2001] {Khan01} Khan, E., Suomij\"arvi, T., Blumenfeld,
Y., et al. 2001, Nucl. Phys. A, 694, 103
\bibitem [Koehler et al., 1998] {ncap06} Koehler, P. E., Spencer, R. R., Guber,
K. H., et al. 1998, Phys. Rev. C, 57, 1158
\bibitem [Koehler et al., 2000] {ncap11} Koehler, P. E., Winters, R. R., Guber,
K. H., et al. 2000, Phys. Rev. C, 62, 055803
\bibitem [Koehler et al., 2001] {ncap13} Koehler, P. E., Harvey, J. A., Winters,
R. R., et al. 2001, Phys. Rev. C, 64, 065802
\bibitem [Koning et al., 2002] {Koning02} Koning, A. J., Beijers, H., Benlliure,
J., et al. 2002, in Proc. of Nuclear Data for Science \& Technology, J. Nucl.
Science \& Technology, Suppl. 2, ed. K. Shibata (Atomic Energy Society of
Japan), 2, 1161
\bibitem [Koning \& Delaroche, 2002]{koning03} Koning, A. J., Delaroche  J.-P. 2003, Nucl. Phys. A, 713, 231
\bibitem [Koning \& Duijvestijn, 2004] {Koning04} Koning, A. J., \& Duijvestijn,
M. C. 2004, Nucl. Phys. A, 744, 15
\bibitem [Koning et al., 2004] {TALYS} Koning, A. J., Hilaire, S., \&
Duijvestijn, M. 2004, TALYS: a nuclear reaction program, NRG-report
21297/04.62741/P, also available at: http://www.talys.eu
\bibitem [Langanke \& Martinez-Pinedo, 2000] {Langanke00} Langanke, K., \&
Martinez-Pinedo, G. 2000, Nucl. Phys. A, 673, 481
\bibitem [Lederer et al., 2011] {ncap57} Lederer, C., Colonna, N., Domingo
Pardo, C., et al. (n-TOF Collaboration) 2011, Phys. Rev. C, 83, 034608
\bibitem [Marrone et al., 2006] {ncap27} Marrone, S., Abbondanno, U., Aerts, G.,
et al. (n-TOF Collaboration) 2006, Phys. Rev. C, 73, 034604
\bibitem [Marganiec et al., 2009a] {ncap44} Marganiec, J., Dillmann, I., Domingo
Pardo, C., et al. 2009a, Phys. Rev. C, 79, 065802
\bibitem [Marganiec et al., 2009b] {ncap46} Marganiec, J., Dillmann, I., Domingo
Pardo, C., \& K\"{a}ppeler, F. 2009b, Phys. Rev. C, 80, 025804
\bibitem [Marganiec et al., 2010] {ncap55} Marganiec, J., Dillmann, I., Domingo
Pardo, C., K\"{a}ppeler, F., \& Walter, S. 2010, Phys. Rev. C, 82, 035806
\bibitem [Meierhofer et al., 2010] {ncap48} Meierhofer, G., Grabmayr, P.,
Jochum, J., et al. 2010, Phys. Rev. C, 81, 027603
\bibitem [M\"oller et al., 1996] {Moller96} M\"oller, P., Nix, J. R., \& Kratz, K.
1996, ArXiv Nuclear Theory e-prints, LA-UR-94-3898
\bibitem [Mohr et al., 1999] {ncap07} Mohr, P., Sedyshev, P. V., Beer, H., et
al. 1999, Phys. Rev. C, 59, 3410
\bibitem [Mosconi et al., 2010] {ncap56} Mosconi, M., Fujii, K., Mengoni, A., et
al. (n-TOF Collaboration) 2010, Phys. Rev. C, 82, 015802
\bibitem [Nakamura et al., 2003] {ncap03} Nakamura, S., Wada, H., Shcherbakov,
O., et al. 2003, J. of Nucl. Science \& Technology, 40, 119
\bibitem [Noguere et al., 2006] {ncap31} Noguere, G., Bouland, O., Brusegan, A.,
et al. 2006, Phys. Rev. C, 74, 054602
\bibitem [O-Brien et al., 2003] {ncap19} O-Brien, S., Dababneh, S., Heil, M., et
al. 2003, Phys. Rev. C, 68, 035801
\bibitem [Oda et al., 1994] {Oda94} Oda, T., Hino, M., Muto, K., Takahara, M.,
\& Sato, K. 1994, Atomic Data \& Nuclear Data Tables, 56, 231
\bibitem [Patronis et al., 2004] {ncap20} Patronis, N., Dababneh, S.,
Assimakopoulos, P. A., et al. 2004, Phys. Rev. C, 69, 025803
\bibitem [Rapp et al., 2002] {ncap16} Rapp, W., Heil, M., Hentschel, D., et al.
2002, Phys. Rev. C, 66, 015803
\bibitem [Ratzel et al., 2004] {ncap22} Ratzel, U., Arlandini, C., K\"{a}ppeler,
F., et al. 2004, Phys. Rev. C, 70, 065803
\bibitem [Reifarth et al., 2002] {ncap18} Reifarth, R., Heil, M., K\"{a}ppeler,
F., et al. 2002, Phys. Rev. C, 66, 064603
\bibitem [Reifarth et al., 2008] {ncap38} Reifarth, R., Heil, M., Forssen, C.,
et al. 2008, Phys. Rev. C, 77, 015804
\bibitem [Samyn et al., 2001] {pro2} Samyn, M., Goriely, S., Heenen, P. H.,
Pearson, J. M., \& Tondeur, F. 2001, Nucl. Phys. A, 700, 142
\bibitem [Samyn et al., 2004] {pro5} Samyn, M., Goriely, S., Bender, M., \&
Pearson, J. M. 2004, Phys. Rev. C, 70, 044309
\bibitem [Sedyshev et al., 1999] {ncap09} Sedyshev, P. V., Mohr, P., Beer, H.,
et al. 1999, Phys. Rev. C, 60, 054613
\bibitem [Shcherbakov et al., 2005] {ncap05} Shcherbakov, O., Furutaka, K.,
Nakamura, S., et al. 2005, J. of Nucl. Science \& Technology, 42, 135
\bibitem [Sin et al., 2006] {WKB} Sin, M., Capote, R., Ventura, A., Herman, M.,
\& Oblozinsky, P. 2006, Phys. Rev. C, 74, 014608
\bibitem [Tachibana et al., 1990] {Tachibana90} Tachibana, T., Yamada, M., \&
Yoshida, Y. 1990, Progress of Theoretical Physics, 84, 641
\bibitem [Tagliente et al., 2008a] {ncap41} Tagliente, G., Fujii, K., Milazzo,
P. M., et al. (n-TOF Collaboration) 2008a, Phys. Rev. C, 77, 035802
\bibitem [Tagliente et al., 2008b] {ncap43} Tagliente, G., Milazzo, P. M.,
Fujii, K., et al. (n-TOF Collaboration) 2008b, Phys. Rev. C, 78, 045804
\bibitem [Tagliente et al., 2010] {ncap49} Tagliente, G., Milazzo, P. M., Fujii,
K., et al. (n-TOF Collaboration) 2010, Phys. Rev. C, 81, 055801
\bibitem [Tagliente et al., 2011] {ncap54} Tagliente, G., Milazzo, P. M., Fujii,
K., et al. (n-TOF Collaboration) 2011, Phys. Rev. C, 84, 015801
\bibitem [Takahashi \& Yokoi, 1987] {Takahashi87} Takahashi, K., \& Yokoi, K.
1987, Atomic Data \& Nuclear Data Tables, 36, 375
\bibitem [Terlizzi et al., 2007] {ncap35} Terlizzi, R., Abbondanno, U., Aerts,
G., et al. (n-TOF Collaboration) 2007, Phys. Rev. C, 75, 035807
\bibitem [Tondeur et al., 2000] {pro1} Tondeur, F., Goriely, S., Pearson, J. M.,
\& Onsi, M. 2000, Phys. Rev. C, 62, 024308
\bibitem [Uberseder et al., 2009] {ncap52} Uberseder, E., Reifarth, R.,
Schumann, D., et al. 2009, Phys. Rev. Lett., 102, 151101
\bibitem [Vockenhuber et al., 2006] {ncap33} Vockenhuber, C., Dillmann, I.,
Heil, M., et al. 2006, Phys. Rev. C, 75, 015804
\bibitem [Voss et al., 1999] {ncap08} Voss, F., Wisshak, K., Arlandini, C., et
al. 1999, Phys. Rev. C, 59, 1154
\bibitem [Winckler, 2006] {ncap01} Winckler, N., Dababneh, S., Heil, M., et al.
2006, ApJ, 647, 685
\bibitem [Wisshak et al., 2000] {ncap10} Wisshak, K., Voss, F., Arlandini, C.,
et al. 2000, Phys. Rev. C, 61, 065801
\bibitem [Wisshak et al., 2001] {ncap50} Wisshak, K., Voss, F., Arlandini, C.,
et al. 2001, Phys. Rev. Lett., 87, 251102
\bibitem [Wisshak et al., 2002] {ncap17} Wisshak, K., Voss, F., K\"{a}ppeler,
F., Kazakov, L. 2002, Phys. Rev. C, 66, 025801
\bibitem [Wisshak et al., 2004] {ncap21} Wisshak, K., Voss, F., Arlandini, C.,
et al. 2004, Phys. Rev. C, 69, 055801
\bibitem [Wisshak et al., 2006a] {ncap24} Wisshak, K., Voss, F., K\"{a}ppeler,
F., et al. 2006a, Phys. Rev. C, 73, 015802
\bibitem [Wisshak et al., 2006b] {ncap26} Wisshak, K., Voss, F., K\"{a}ppeler,
F., Kazakov, L. 2006b, Phys. Rev. C, 73, 015807
\bibitem [Wisshak et al., 2006c] {ncap28} Wisshak, K., Voss, F., K\"{a}ppeler,
F., et al. 2006c, Phys. Rev. C, 73, 045807
\bibitem [Xu et al., 2009] {nacre2pro4} Xu, Y., Goriely, S., Takahashi, K., in: Susa, H., Utsunomiya, H., Arnould, M., Gal\`es, S., Motobayashi, T., Scheidenberger, C. (Eds.), Proceedings of Tours Symposium on Nuclear Physics VII, Kobe, Jaoan, November 16-20, 2009, AIP Conf. Proc., 1238, 187-192
\bibitem [Xu et al., 2012] {NACRE2} Xu, Y., Takahashi, K., Goriely, S., Arnould, M., Ohta, M., \& Utsunomiya, H. 2012, submitted to Nucl. Phys. A (NACRE II)

\end{thebibliography}
\end{document}